\documentclass{article}
\usepackage[american]{babel}
\usepackage{amsopn}
\usepackage{amsfonts}
\usepackage{hyperref}
\usepackage{amsmath,amssymb}
\usepackage[dvips]{graphicx}
\usepackage{eurosym}
\usepackage{float}
\numberwithin{equation}{section}
\def\xE{{\mathbb E}}
\def\xLone{{\rm L}^1}
\def\xR{{\mathbb R}}

\newtheorem{thrm}{Theorem}[section]

\newtheorem{rmrk}[thrm]{Remark}

\begin{document}
\title{Bayesian Estimation of Inequalities with Non-Rectangular Censored Survey Data}

\author{Eric Gautier}

\date{}

\maketitle
\begin{center}
ENSAE - CREST, 3 avenue Pierre Larousse, 92245 Malakoff Cedex,
France; \texttt{gautier@ensae.fr}
\end{center}

\begin{abstract}
Synthetic indices are used in Economics to measure various aspects of monetary inequalities.
These scalar indices take as input the distribution over a finite population, for example the
population of a specific country. In this article we consider the case of the French 2004 Wealth survey.
We have at hand a partial measurement on the distribution of interest consisting of bracketed and sometimes
missing data, over a subsample of the population of interest. We present in this article the statistical
methodology used to obtain point and interval estimates taking into account the various uncertainties.
The inequality indices being nonlinear in the input distribution, we rely on a simulation based
approach where the model for the wealth per household is multivariate. Using the survey data as
well as matched auxiliary tax declarations data, we have at hand a quite intricate non-rectangle
multidimensional censoring. For practical issues we use a Bayesian approach. Inference using
Monte-Carlo approximations relies on a Monte-Carlo Markov chain algorithm namely the Gibbs
sampler. The quantities interesting to the decision maker are taken to be the various inequality
indices for the French population. Their distribution conditional on the data of the subsample
are assumed to be normal centered on the design-based estimates with variance computed through
linearization and taking into account the sample design and total nonresponse. Exogeneous
selection of the subsample, in particular the nonresponse mechanism, is assumed and we condition
on the adequate covariates.\vspace{0.3cm}

\vspace{0.3cm}

\end{abstract}

\section{Introduction}\label{s1}
Approximately every six years the French statistical office INSEE collects a cross sectional
wealth survey on households. The last dataset was collected in 2004. Several aspects can be
studied focusing for example on holdings or particular types of assets like
the professional wealth or intergenerational transfers. One natural question concerns the nature
of the distribution of wealth and its allocation in the various possible holdings.
However, it is known to be a difficult one as well (Juster and Smith 1997) due to the
difficulty to have good measurements and to possible selection biases. Questions on income
and wealth are particularly sensitive and the nonresponse probability is very likely to be
related to the value itself, resulting in possible endogeneous selection and biases. Also,
it is for example particularly difficult to give a precise amount for the market value of one's
real estate piece of property unless people assessed it recently, say in order to sell it.
Thus, amounts are usually collected in bracketed format and imputation methods (see experiment
based on the knowledge of the true income distribution in Lollivier and Verger 1988) are used
in practice at the French institute.\\
\indent For some variables the brackets are defined by each household,
they give upper and lower bounds for the amount based on their evaluation. For other variables the
households choose among a predefined system of brackets. The method allows to replace missing data,
impairing inference due to selection bias and loss of efficiency implied by a reduced sample, by
censored data. In the current article we focus on the evaluation of inequality indices on the total
wealth for the whole French population. A specific
question at the end of the survey: "Suppose you had to sell everything, how much do you assess the
value of your total wealth including durable goods, artwork, private collections and jewelry" allows to
measure the total wealth.
The values of the last items were not collected. It is troublesome to ask such information because the pollster
comes to the household's home and they could be suspected of theft in case a robbery
occurs after the visit. The system of brackets for the question collecting the total wealth
has an unbounded last bracket. The threshold for the higher bracket is 450.000 \euro{} which is pretty low.
Also in order to improve, in principle, the precision of design based estimates, certain categories have been
over-sampled: self employed, executives, retired people and people living in rich neighborhoods.
These variables,
available from the census, are indeed correlated with the wealth and sampling more a priori wealthy people
improve the precision of design based estimates of inequality indices sensitive to the top of the distribution.
But due to the censoring, a billionaire is equivalent to an household whose total wealth is 451.000 \euro{}.
Over-sampling has increased the number of household for which we measure wealth very imprecisely.
Thus, though we are interested in the particular concept of total wealth collected in this final question,
we aim to gather more information in order to better estimate the inequality indices.
Due to the high amount of censoring for the wealthiest, the less wealthy contribute the most to the
likelihood. Moreover, some of the indices are quite sensitive to misspecification of the model.
Though it is pretty usual for wages and even for wealth to specify linear models for the logarithm
with normals residuals, some of the assumptions like the distributional assumption for the residuals are not
testable in the absence of pointwise measurements. Other distributions like Pareto are also quite popular in
the literature on wealth inequalities (Lollivier and Verger 1988). It is also possible that the
influence of certain covariates is not additive, for example the contribution
of the income might differ for low incomes from that of high incomes, and, due to the
censoring of high wealth, it is possible that we do not capture well this mixture
(see for example Duclos et al. 2004 and Esteban and Ray 1994 for theoretical foundations
of polarization). Another possible source of misspecification is heteroscedasticity. If for example
the variance of the residuals increase with income, censoring and specification of an homoscedastic
model will imply lower inequality indices.
Therefore, we gather more information in order to recover better knowledge of
the top of the distribution of wealth and better estimate the total population indices.
We use bracketed information on components of the total wealth as well as bracketed
information involving several components: the total wealth (sum of the components) and the
information on the imposition on the Solidarity Tax on Wealth (ISF) obtained by matching with data
from the tax declarations.\\
\indent Our approach relies on Bayesian multiple imputations (Little and Rubin 2002) for sample survey
estimation. However, our strategy to produce point estimates or interval estimates is slightly
different. We do not rely on particular rules for combining complete-data inferences or rely
on imputations which are Bayesianly proper (Schafer 2001). We use a hierarchical modeling
in order to take into account in the coverage of the interval estimates the uncertainty due
to sampling and total-nonresponse and to incomplete knowledge of the censored wealth. Under
the assumption that the proper parametric class for the data generating process (DGP) is known,
the uncertainty on the value of the censored wealth boils down to uncertainty on the parameters
values and to the remaining model uncertainty due to imperfect observation conditional on the knowledge
of the parameters. The
first model is the model for the quantities interesting to the decision maker (Geweke 2005)
which are here taken to be the various inequality indices on the finite French population.
The remaining models, which are standard, are instrumental in the sense that it is not our
goal to produce inference on the posterior predictive distribution of the wealth of the
sampled households or on the posterior distribution of the parameters, though they could be obtained
simultaneously with the numerical procedure. In order to gather information
on the components of total wealth, our DGP is multivariate. It allows to account for
example for unobserved heterogeneity.
From that point of view, our approach is similar to that of Heeringa et al. (2002),
where a multivariate model is used in the context of the American Health and Retirement
Survey (HRS). Inference is based on a single path of a Gibbs sampler Markov chain that
updates the sampled data, parameters and an error term accounting for sampling error.
The only mathematical tool is the Ergodic theorem. The model is discussed in Section
\ref{s2}, the non-rectangular censoring is presented in Section \ref{s3}. In Section \ref{s4}
we give details on the Gibbs sampler and discuss our strategy to produce point and interval estimates.
Results are presented in Section \ref{s5} and it is followed by a discussion in Section \ref{s6}.

\section{The Modeling Assumptions}\label{s2}
We denote by $U$ the finite population of size $N$ composed of all the French households
and $S\subset U$ the sample, where we number the elements of $S$ from $1$ to $m$. Due to
total nonresponse, $S$ corresponds to a subset of the initial sample drawn from particular
bases of dwellings. The initial sample is stratified and drawn with unequal probabilities
where dwellings composed, at the time of the census, of self employed, executives,
retired people and of people living in rich neighborhoods have been over-sampled. It implies
that probability of selection is related to the wealth but, in principle, the selection is exogeneous.
We assume below that the selection mechanism corresponding to the total nonresponse is also exogeneous
and that in the models we have included the adequate covariates to be able to ignore the selection
mechanism (Little and Rubin 2002).\\
\indent The target quantities or quantities of interest are taken to be inequality indices
on the total wealth of the French: the Gini,
Theil or Atkinson's indices, quantiles or inter-quantile ratios.
They are functions of the distribution of the total wealth $t_k$ for
households $k$ from 1 to $N$. Recall that, for example, the Gini is defined by
$$G=\frac{\sum_{k\in U}(2r(k)-1)t_k}{N\sum_{k\in U}t_k}-1,$$
where $r(k)$ is the rank of $t_k$. A design based estimate is then
$$\hat{G}=\frac{\sum_{k\in U}(2\hat{r}(k)-1)w_kt_k}{\sum_{k\in S}w_k\sum_{k\in S}w_kt_k}-1,$$
where $w_k$ is the weight of household $k$,
$\hat{r}(k)=\sum_{j\in S}w_jI\left\{t_j\le t_k\right\}$ and $I\{\cdot\}$ denotes
the indicator function.
In practice, at INSEE, a normal approximation for the design based estimate
is used in order to obtain interval estimates. Also, since the variance of the estimate requires
in principle as well the data in $U\setminus S$, a variance estimate is used.
$\hat{G}$ being nonlinear in the weights, the estimate of the variance of $\hat{G}$
is approximated by that of its linearized version. The variance estimate should
take into account the complex design of the sample and the total nonresponse and
raking (Deville et al. 1993). The procedure is well explained in Dell et al.
2002. It is however difficult to justify all the approximations rigorously. We do
not aim to enter in these details and start off from the approximation:
$$\hat{G}\approx G+\sqrt{\widehat{V\left(\hat{G}\right)}}E$$
where the error term $E$ is a standard centered Gaussian random variable and the
variance estimate is denoted by $\widehat{V\left(\hat{G}\right)}$.\\
\indent Since the total wealth is censored, we are not able to apply the above tools to estimate the target quantities.
We rely a priori on a two stage model. But, since for practical issues we have adopted
the Bayesian point of view, we have added an additional ladder to the hierarchy of models:
\begin{enumerate}
\item The model (I) for the quantities of interest like the Gini, conditional on the relevant data
from the sampled household.
\item The model (DGP) for the components of the wealth for sampled households. It is
a multivariate model for owned macro-components of the total wealth among: the financial
wealth $W^1$, the value of the principal dwelling $W^2$, of the
other real estate including secondary dwellings rented or for leisure and parking lots $W^3$, the
professional wealth $W^4$ and the remainder (durable goods, artwork, private collections and jewelry) $W^5$,
conditional on the value of covariates ${\bf x_{k}^{l}}$, $k=1,\hdots m$, $l=1,\hdots,5$ and on the
parameters in a certain parametric class of models.
\item The prior distribution (P) of the parameters ${\bf \Theta}$ of density $\pi(\theta)$.
\end{enumerate}
For simplicity, we assume that every household has some financial wealth (eg. money on a checking account)
and some wealth in form of remainder. Therefore, if we make groups according to the type of portfolio in the 5
above components of wealth we have to distinguish 8 groups. We denote by ${\bf D}_k=(D_k^l)_{l=1,\hdots,5}$
the binary vector such that $D_k^l=I\left\{W_k^l>0\right\}$ and define the function $P$ which associates to each $\bf{D}_k$
the number $i\in{1,\hdots,8}$ of the pattern. In the remaining, we use capital letters for random variables and lowercase
letters for realizations. We also use bold characters for vectors.\\
\indent The model is defined as follows. In the first stage (I) we set, for example for the Gini,
\begin{equation}\label{eI}
G=\hat{G}\left(t_1,\hdots,t_m\right)+\sqrt{\widehat{V\left(\hat{G}\right)}}
\left(t_1,\hdots,t_m\right)E,\quad E\rightsquigarrow
\mathcal{N}(0,1)
\end{equation}
with the Assumption (A):
\begin{center}
$E$ independent of $\left(t_1,\hdots,t_m\right)$\hspace{0.5cm}(A).
\end{center}
Concerning the model (DGP), we have the following model for pattern $i$:
\begin{equation}\label{eDGP}
\left\{\begin{array}{l}
        T_k=\sum_{l=1}^5W_k^l,\\
        \log(W_k^l)={\bf x_{k}^{l}}{\bf \beta_i^l}+U_{i}^{l},\ {\rm when}\ d_k^l=1\ {\rm and}\
        P\left({\bf d}_k\right)=i\\
        {\bf U_i}\rightsquigarrow\mathcal{N}(0,\Sigma_i)
      \end{array}\right.
\end{equation}
where $\Sigma_i$ is of size $p_i=\sum_{l=1}^5d_k^l$ for any $k$ such that
$P\left({\bf d}_k\right)=i$.
We make the following restriction on the parameters:
\begin{center}
Only the coefficient of the constant is group specific (fixed effect), the remaining
coefficients of ${\bf \beta_i^l}$ are equal for all $i$\hspace{0.5cm} (RP).
\end{center}
For the last model (P), we choose $\pi(\theta)$
proportional to
\begin{equation}\label{eP}
\prod_{i=1}^8\det\left(\Sigma_i\right)^{-\frac{p_i+1}{2}}.
\end{equation}
In usual design based inference in survey sampling, the Gini index $G$ has an unknown
but fixed value. Hence, Equation \eqref{eI} is not usual since $G$ is now random. In some
approaches to survey sampling though, the finite population values correspond to draws
in a super-population and it makes sense to assume that the quantities of interest
on the finite population are random. It is also usual to revert the Gaussian approximation
to obtain interval estimates. We may also think about estimating $G$ in terms of prediction.\\
\indent We present in Table \ref{Ta1} the kind of covariates we have introduced in the model (DGP).
\begin{table}[H]
  \centering
  \caption{Covariates for the model (DGP) other than the type of portfolio }\label{Ta1}
{\small
  \begin{tabular}{|l|c|c|c|c|c|}
  \hline
  {\bf Covariate / Component} & $W^1$ & $W^2$ & $W^3$ & $W^4$ & $W^5$ \\
  \hline
  {\bf Life cycle} &  &  &  &  &  \\
  \hspace{0.1cm} single and no child &  & $\surd$ & $\surd$ & $\surd$ & $\surd$ \\
  \hspace{0.1cm} age and age square &  & $\surd$ & $\surd$ & $\surd$ & $\surd$ \\
  \hspace{0.1cm} position in the life cycle\footnote{detailed variable which interacts age, number of children and family type} & $\surd$ &  &  &  & \\
  \hline
  {\bf Social and Education} &  &  &  &  &  \\
  \hspace{0.1cm} social/professional characteristics & $\surd$ & $\surd$ & $\surd$ & $\surd$ & $\surd$ \\
  \hspace{0.1cm} higher educational degree & $\surd$ & $\surd$ & $\surd$ & $\surd$ & $\surd$ \\
  \hline
  {\bf Income} &  &  &  &  &  \\
  \hspace{0.1cm} level of the salary & $\surd$ & $\surd$ & $\surd$ & $\surd$ &  \\
  \hspace{0.1cm} social benefits received & $\surd$ &  &  &  &  \\
  \hspace{0.1cm} rent received &$\surd$ & $\surd$ &  & $\surd$ &  \\
  \hspace{0.1cm} other income received & $\surd$ &  & $\surd$ & $\surd$ &  \\
  {\bf Location of the residence} & $\surd$ & $\surd$ & $\surd$ &  & $\surd$ \\
  {\bf History of the wealth} &  &  &  &  &  \\
  \hspace{0.1cm} donation received & $\surd$ & $\surd$ &  & $\surd$ & $\surd$ \\
  \hspace{0.1cm} donation given & $\surd$ &  &  &  &  \\
  \hspace{0.1cm} recent increase/decrease of wealth & $\surd$ & $\surd$ &  & $\surd$ & $\surd$ \\
  \hspace{0.1cm} type of wealth of the parents & $\surd$ &  & $\surd$ & $\surd$ &  \\
  {\bf Surface and square of the surface} &  & $\surd$ &  &  &  \\
  {\bf Professional wealth} &  &  &  &  &  \\
  \hspace{0.1cm} wealth used professionally &  &  &  & $\surd$ &  \\
  \hspace{0.1cm} firm owned &  &  &  & $\surd$ &  \\
  \hline
\end{tabular}}
\end{table}
Covariates can also improve a priori
the coverage of the interval estimates, up to a certain stage since increasing the size of the vector of
parameters deteriorates the knowledge on the parameters. The main justification for introducing the covariates
is however to justify Assumption (A). Indeed, the survey sample is drawn exogeneously and we have to condition
by the corresponding observed covariates in order to estimate the law of the data unconditional
of the selection. Total nonresponse is a second stage of selection for which the selection mechanism is
unknown and we assume that this mechanism is ignorable (Little and Rubin 2002 and Gautier 2005) and that
we condition by the adequate covariates to decondition from selection.\\
\indent The model (DGP), is such that, though we take into account observed
heterogeneity in the form of portfolio allocation and through several covariates, there might remain
unobserved heterogeneity (eg. in the form of a missing covariate) like the preference for the risk and time
that causes the residuals to be dependent. In order to use product specific variables for the principal dwelling
we model the value of the good. In contrast in the other models, for which the variables are sums of components collected in the survey,
we model the amount of the share
that the household possesses and use household specific variables only.\\
\indent The vector of parameters ${\bf \theta}$ in $\xR^d$ corresponds to the ${\bf \beta_i^l}$'s and the matrices
$\Sigma_i$ where, denoting by ${\rm dim}_l$ the dimension of ${\bf \beta_{i,l}}$,
$$d=\sum_{l=1}^5\left({\rm dim}_l-1\right)+8*5+\frac12\sum_{k=2}^5k(k+1).$$
The prior is a product of usual priors in the context of Gaussian linear models which are limits
of normal/inverse-Wishart's (see for example Little et Rubin 2002 and Schafer 2001).
They are often called non-informative. They indeed correspond to a proper objective choice for the
prior for the coefficients ${\bf \beta_i^l}$. The posterior, if the data were observed, is a {\it bona-fide}
normal/inverse-Wishart probability distribution.

\section{Censoring}\label{s3}
In the absence of scalar measurements, intervals are the main information
for identification and estimation. As already discussed, we aim to localize as much as possible
the missing ${\bf W_k}$. For that purpose we use two summarizing questions and the information
on which household is eligible to the Solidarity Tax on Wealth.
The answers to the summarizing questions take the form of brackets for the sum of the collected
components of the financial wealth as well as brackets for the total wealth.
Recall that the total wealth includes the
remainder which is not collected {\it per se} in the detailed questionnaire. The information on which
household pays the Solidarity Tax on Wealth has been obtained by matching with a data set from the
tax department.The condition to pay the Solidarity Tax on Wealth is to have a taxable wealth exceeding 720.000
\euro{}. This taxable wealth corresponds to a different concept of wealth.
Only part of the professional wealth is taken into account. It is possible to deduct the
professional wealth used professionally with the exception that if one owns a
share in a firm which is too low then it is not deductible.
It is possible to have a rebate of
20\% on the value of one's principal dwelling. The artworks are not taxed either. Finally, debts are
deducted. It is possible to take into account most of the specificities of this tax.
For example, by chance, the few households that possessed a share in a firm gave equal upper
and lower bounds for its value. However, it is not possible to distinguish the artworks within the remainder.
Hence, we produce lower and upper bounds on the taxable wealth which take the form
of an extra bracketing condition.\\
\indent When an household pays the tax, the upper bound for the taxable wealth
\begin{equation}
W_k^1+0.8*W_k^2+W_k^3+\min(W_k^4,NDED_{\max,k})+W_k^5-DEBT_k
\end{equation}
is greater than 720.000 \euro{}, where $NDED_{\max,k}$ is an upper bound of the nondeductible professional
wealth obtained using the detailed information and $DEBT_k$ the total debts which is deductible.
We assume that households always subtract the deductible amounts.\\
\indent When an household does not pay the tax, the lower bound for the taxable wealth
\begin{equation}
W_k^1+0.8*W_k^2+W_k^3+NDED_{\min,k}-DEBT_k
\end{equation}
is lower than 720.000 \euro{}, where $NDED_{\min,k}$ is a lower bound of the nondeductible
professional wealth obtained using the detailed information.\\
\indent The above conditions involving several variables allow, by manipulation of lower and upper bounds,
to shorten the initial intervals for each component and to obtain intervals for the remainder.
The censoring takes form of a hyper-rectangle.
The final summarizing condition for the total wealth
and the eligibility to the Solidarity Tax on Wealth imply censored domains which are subsets of these
hyper-rectangles.\\
\indent Note that further external information has also been used in order to specify upper
bounds on the a priori unbounded total wealth. This choice is questionable but these
upper bounds are very loose. The motivation is that, since design based inference is
the initial goal and each sampled household has a weight that is the inverse of the probability of
selection\footnote{It is in fact estimated due to total nonresponse.} and the data
is collected once and for all, we want to have reasonable "representativity" of the
samples\footnote{The samples are simulated, as explained later}.
The average weight is around 2.000. Suppose a billionaire is drawn, then it is assumed to
represent 2.000 households. In turn, if we were able to draw at random a second time the sample,
it is very likely that no billionaire would be drawn. It might result in inequalities which are often
too low and sometimes too high. Probably due to the over-sampling of households suspected to be wealthy,
an household with a share in a firm of the order of 25.000.000 \euro{} has been drawn.
We have introduced upper bounds on the total wealth based on Cordier et al. (2006) and published information
on the highest French professional wealth.
We have bounded by 50.000.000 \euro{} the total wealth of the apparently wealthiest household and by
10.000.000 \euro{} the total wealth of the others. Such restriction might cause under coverage
of the interval estimates. Heeringa, Little and Raghunatan (2002) have noted, using a similar modeling and the HRS survey,
that introducing such restrictions implies slightly lower means, Ginis or other concentration indices.
On the other hand its impact on more robust quantities such as quantiles is minor.

\section{Numerical Procedure for Point and Interval Estimates}\label{s4}
\subsection{The Gibbs sampler}
We adopt the Bayesian point of view for practical reasons.
From a frequentist point of view, the first stage consists in the estimation
of the parameters of the multivariate linear Gaussian model with censored
observations. Several methods are at hand among which  simulated based methods
like the simulated maximum likelihood, simulated scores or simulated method of moments
(see Train 2003 for a lively and basic introduction) or a MC-ECM variant of the EM
algorithm (Little and Rubin 2002).
Also, since the criterion function has in general several local extrema, it
is often useful to use a stochastic optimization method or at least to try several
initial starting points. It is also useful, as we will see below, to simulate in the
multivariate truncated normals in order to finally infer on the finite population
inequality indices. The GHK simulator (Geweke 2005) in the non-rectangular context
is not standard and also requires proper importance sampling weighting which in the
context of our hierarchical modeling does not seem feasible. Accept-reject with
instrumental distribution the unconditional distribution is known to be very ineffective,
especially as the dimension increases. Robert (1995) for example suggests the use of the
Gibbs sampler (Arnold 1993). The total procedure is extremely intensive from a
computational viewpoint. Moreover, we need to modify slightly the procedure
if we want to take into account the uncertainty on the parameters due to the finite
distance.\\
\indent The Gibbs sampler easily adapts to the Bayesian modeling, see for example
McCulloch and Rossi (1994) for Bayesian inference for multinomial discrete
choice models. It is also popular in missing data problems (Little and Rubin
2002 and Schafer 2001) and the extension is called data augmentation. The
only difference with the usual Gibbs sampler is that the state space of the
Markov Chain is augmented in order to include the parameters. Also, in order
to infer on the quantities of interest, we augment the state space once more
and include the $E$'s. The Gibbs sampler relies on an exhaustive block
decomposition of the coordinates of the state space. These blocks are numerated
according to a specific order. Starting from an initial value ${\bf v_0}$, the
Gibbs sampler simulates a path from a Markov chain $\left({\bf v_n}\right)_{n\ge0}$.
Given ${\bf v_n}$, a vector ${\bf V_{n+1}}$ decomposed in the above system of blocks
is simulated by iteratively updating the blocks and sampling from the distribution
of the block conditional on the values at stage $n$ of the future blocks and the
value at stage $n+1$ of the previously updated blocks. Here ${\bf V_n}$  is taken to
be $${\bf V_n}=\left({\bf \Theta,W_1',\hdots,W_m'},E\right)'.$$ The sequence is such that
we start by updating the covariance matrices, followed by the ${\bf \beta_i^l}$, then by the
components of wealth one by one, household by household and finish with the error term in model
(I). The updating for the distribution without truncation is for example explained in Little and
Rubin (2002). Here, we simulate the components of wealth in truncated univariate normals, which
is easy and efficient. We update the intervals of truncation for the current variable
at each stage of the sequence with the previously simulated components for the same household.\\
\indent The limit theorems for the Gibbs sampler are given in Tierney (2004) and Roberts
and Smith (1994). We can also check as in Roberts and Polson (1994), minorizing the transition
kernel using that we have introduced upper bounds for the a priori unbounded amounts, that there
is uniform exponential $\xLone$ ergodicity. Thus convergence of the laws of the marginals of the Markov chain to
the target joint posterior and posterior predictive and distribution\footnote{Recall that it
is always independent of the rest of the components.} of $E$, which is the invariant
probability $\mu$, is very fast. The main result on Markov chains which is useful
for the inference here is the ergodic theorem. It states that for $g$ in $\xLone(\mu)$,
\begin{equation}\label{eergodic}
\lim_{T\rightarrow\infty}\frac{1}{T}\sum_{n=1}^Tg\left({\bf V_n}\right)=\xE_{\mu}
\left[g\left({\bf V}\right)\right]\quad a.s.
\end{equation}

\subsection{Posterior Predictions and Posterior Regions}
\begin{figure}[H]\centering\label{fig1}
\includegraphics[height=0.4\hsize]{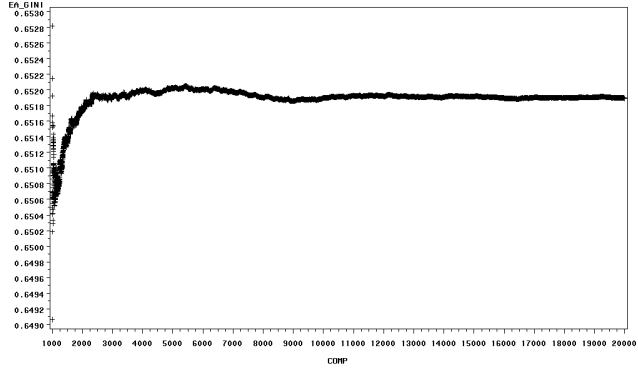}
\caption{Convergence of the empirical averages of the Gini, burn-in of the 1.000 first iterations}
\end{figure}

Suppose that the decision maker wants the statistician to give him a single value for each
quantity of interest.
A natural question is then to ask: "What is the optimal answer one can give?".
Once a loss function, say quadratic, is specified, the optimal answer
\begin{equation}\label{ePred}
\overline{G}=\xE_{\mu}\left[G\left({\bf V}\right)\left|{\bf W_1}\in D_1,\hdots,
{\bf W_m}\in D_m,{\bf X_1}={\bf x_1},\hdots,{\bf X_m}={\bf x_m}
\right.\right],
\end{equation}
among all answers $G^*$, minimizes the posterior risk
\begin{equation}\label{ePR}
\xE_{\mu}\left[\left(G^*-G\left({\bf V}\right)\right)^2\left|{\bf W_1}\in T_1,
\hdots,{\bf W_m}\in T_m,{\bf X_1}={\bf x_1},\hdots,{\bf X_m}={\bf x_m}
\right.\right]
\end{equation}
where the domains $T_k$ correspond to the domains of truncation in $\xR^{p_{P({\bf D_k})}}$
and ${\bf x_k}$ are the matrices of the covariates\footnote{For example block diagonal with
$p_{P({\bf D_k})}$ rows.}. The expectation giving the posterior prediction
in \eqref{ePred} could be approximated by a MCMC method (Robert and Casella 2004): the empirical mean along one
path using \eqref{eergodic}. Since $T$ is chosen by the statistician, this approximation of the
integral could be as good as one wishes. As usual in MCMC methods we present in Section
\ref{s5} results with burn-in, {\it i.e.} where we have dropped the first $B$ simulations.
According to the Cesar\`o lemma, \eqref{eergodic} still holds starting the sum from $n=B+1$ and replacing $T$ by $T-B$.
Heuristically, it allows to wait for for the chain to stabilize close to the steady state.
It only changes the very last decimals here since we have taken $T=20.000$, which is very large as
attested by Figure \ref{fig1}, and $B=1.000$.
With the optimality as a goal, simple imputation is no good. The researcher arbitrarily chooses
one scenario for $G$ among an infinity of possible scenarii. Also, it is natural that predicting
the quantities of interest is different from predicting the unobserved wealth due to the
nonlinearity of $G$ in the wealth. Predicting the unobserved wealth is doomed to introduce biases.\\
\indent Suppose that the statistician has convinced the decision maker that it is better to be given
an interval estimate. We take in Section \ref{s5} symmetric $95\%$ posterior regions but we could
have taken highest posterior regions\footnote{HPD regions.} or posterior regions of minimal length.
The boundaries of the intervals $[l,u]$ can be obtained, without Central Limit Theorem and with a single
path,
by inverting the functional:
\begin{equation}\label{eCR}
\xE_{\mu}\left[I\left\{G\left({\bf V}\right)\in[l,u]^c\right\}\right]\le 0.05
\end{equation}
where the left hand-side could be well approximated for large $T$ by
\begin{equation}\label{eapproxCR}
\lim_{T\rightarrow\infty}\frac1T\sum_{n=1}^TI\left\{G\left({\bf v_n}\right)\in[l,u]^c\right\}
\end{equation}
using \eqref{eergodic}. Again it is possible to use burn-in.
\begin{rmrk}
Unlike the original Bayesian multiple imputations (Little and Rubin 2002 and Schafer
2001) we do not require proper Bayesian imputations, {\it i.e.} independent sampling,
nor rely on formulas to combine multiple imputations. Multiple imputations here is only
a tool to infer on the quantities of interest.
\end{rmrk}
Such interval estimates
take into account the uncertainty due to sampling and total nonresponse, to imprecise
knowledge of the value of the parameters among a parametric class of models due to finite
sample size and to the uncertainty due
to the imperfect measurement of the components of wealth conditional on the knowledge of
the parameters.

\section{Presentation of the Results}\label{s5}
We have applied the methodology of Section \ref{s4} to the data set and runned a Gibbs
sampler with $T=20.000$ and $B=1.000$. We have tried to diagnose convergence by plotting
the convergence of empirical averages required for the inference. As expected due to
exponential ergodicity, the convergence seems to occur very quickly. We have plotted
in Figure 1 the convergence of the empirical averages for the Gini.
\noindent In Table \ref{Ta2} we collect the estimated posterior predictions and confidence regions.
\begin{table}[H]
  \centering
  \caption{Posterior predictions and posterior regions (lower and upper bound of a $95\%$ symmetric region),
  T=20.000, burn-in of the 1.000 first iterations}\label{Ta2}
  \begin{tabular}{|l|c|c|c|}
    \hline
    Quantity of interest & Prediction & Lower bound& Upper bound\\
    \hline
    Mean (\euro{})& 205.003,98 & 192.879,50 & 217.647,24\\
    \hline
    Median (\euro{})& 111.459,26 & 105.672,32 & 117.563,45\\
    \hline
    P99 (\euro{})& 1.584.602,96 & 1.359.261,98 & 1.825.362,43\\
    \hline
    P95 (\euro{})& 690.793,96 & 636.924,50 &	746.759,31\\
    \hline
    P90 (\euro{})& 434.458,13 & 416.410,51 & 452.006,79\\
    \hline
    Q3 (\euro{})& 232.307,50 & 224.849,03 & 240.204,86\\
    \hline
    Q1 (\euro{})& 16.998,67 & 15.117,08 & 19.149,60\\
    \hline
    P10	(\euro{})& 3.959,07 & 2.870,83 & 5.070,55\\
    \hline
    P95/D5	& 6,1972 & 5,7232 & 6,6808\\
    \hline
    P99/D5	& 14,2175 &	12,1667	& 16,4388\\
    \hline
    Q3/Q1	& 13,6847 &	12,2838	& 15,1034\\
    \hline
    D9/D1	& 109,9332 & 80,7615 & 140,5827\\
    \hline
    D9/D5	& 3,8981 & 3,7081 &	4,0858\\
    \hline
    Gini	& 0,6519 & 0,6328 & 0,6717\\
    \hline
    Theil	& 0,9044 &	0,8138 & 1,0001\\
    \hline
    Atkinson ($\epsilon=1.5$) &	0,9063 & 0,8838 & 0,9253\\
    \hline
    Atkinson ($\epsilon=2$)	& 0,9742	& 0,9549	& 0,9920\\
    \hline
  \end{tabular}
\end{table}
\noindent We finally represent in Figure 2 histograms for the posterior distribution of some of the target
quantities of interest.
\begin{figure}[H]\label{fig2}
\includegraphics[height=0.38\hsize]{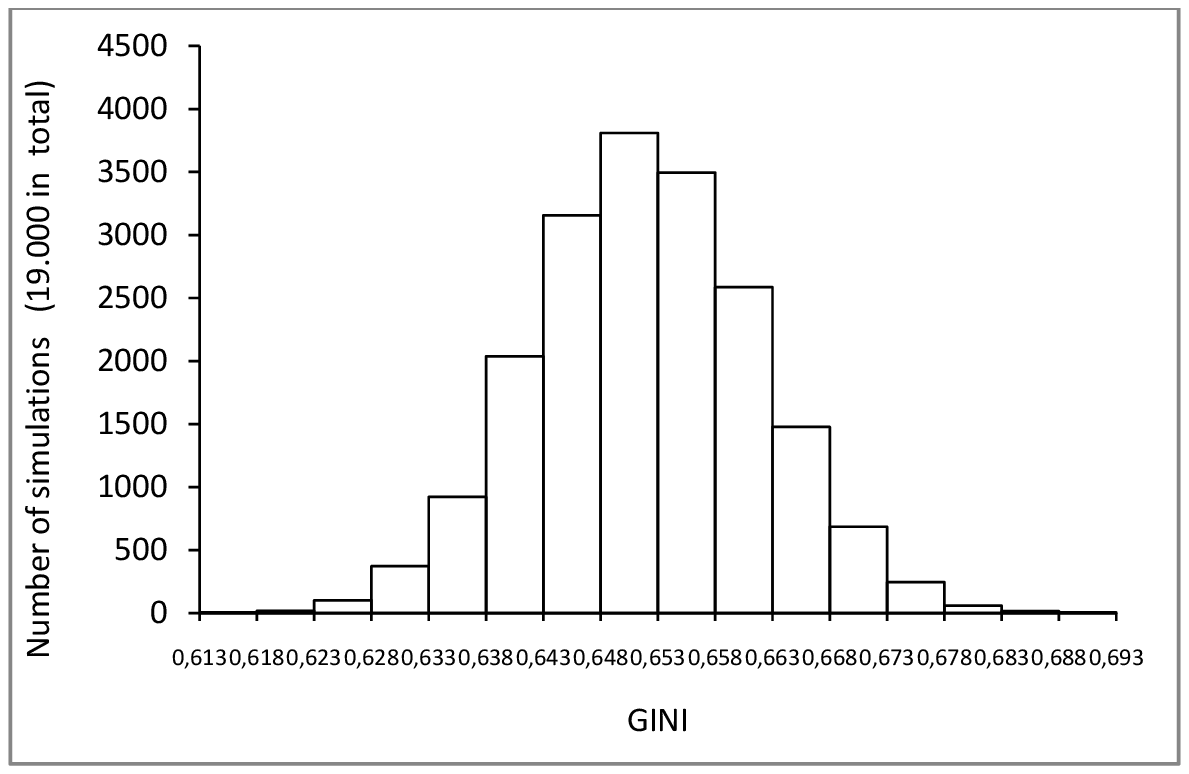}\includegraphics[height=0.37\hsize]{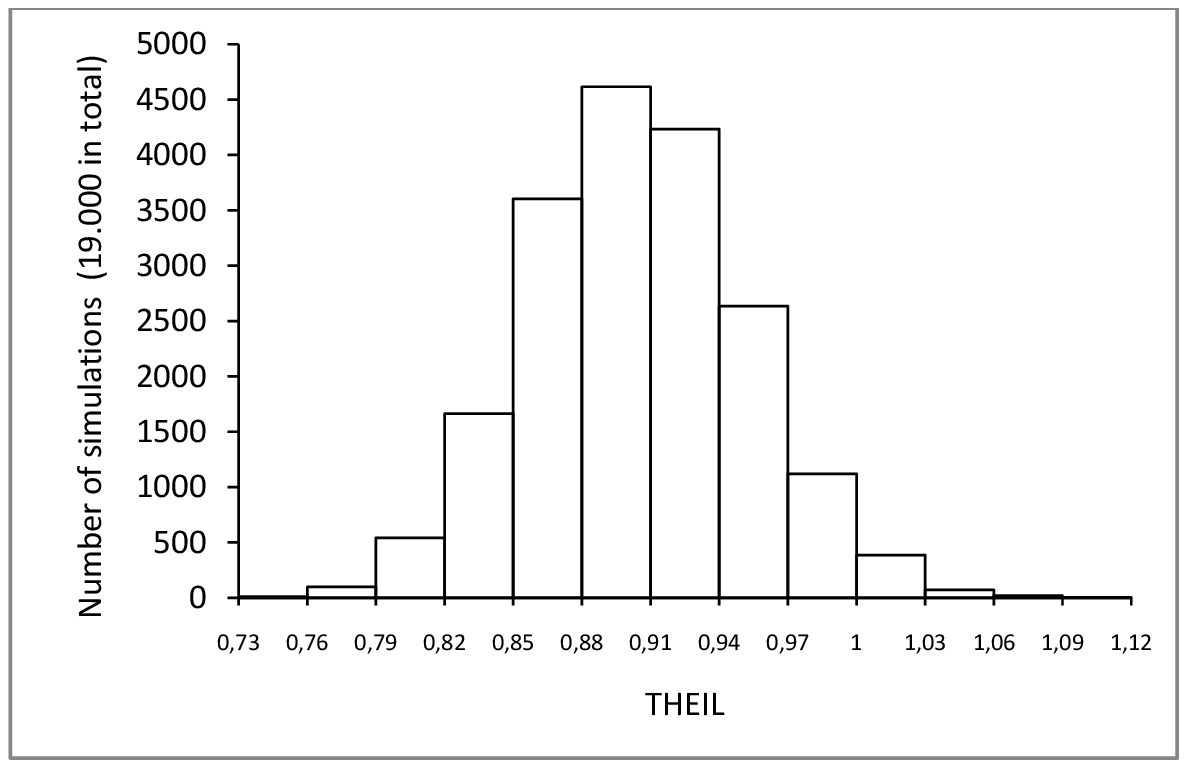}
\includegraphics[height=0.37\hsize]{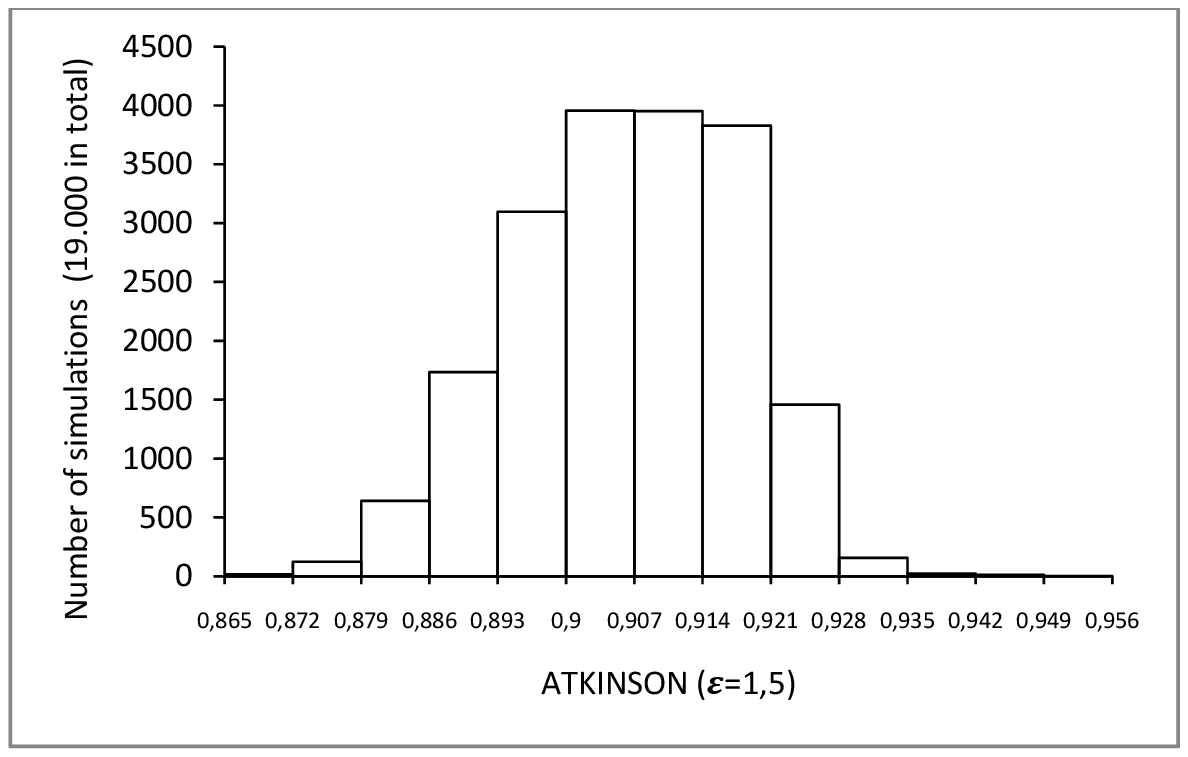}\includegraphics[height=0.37\hsize]{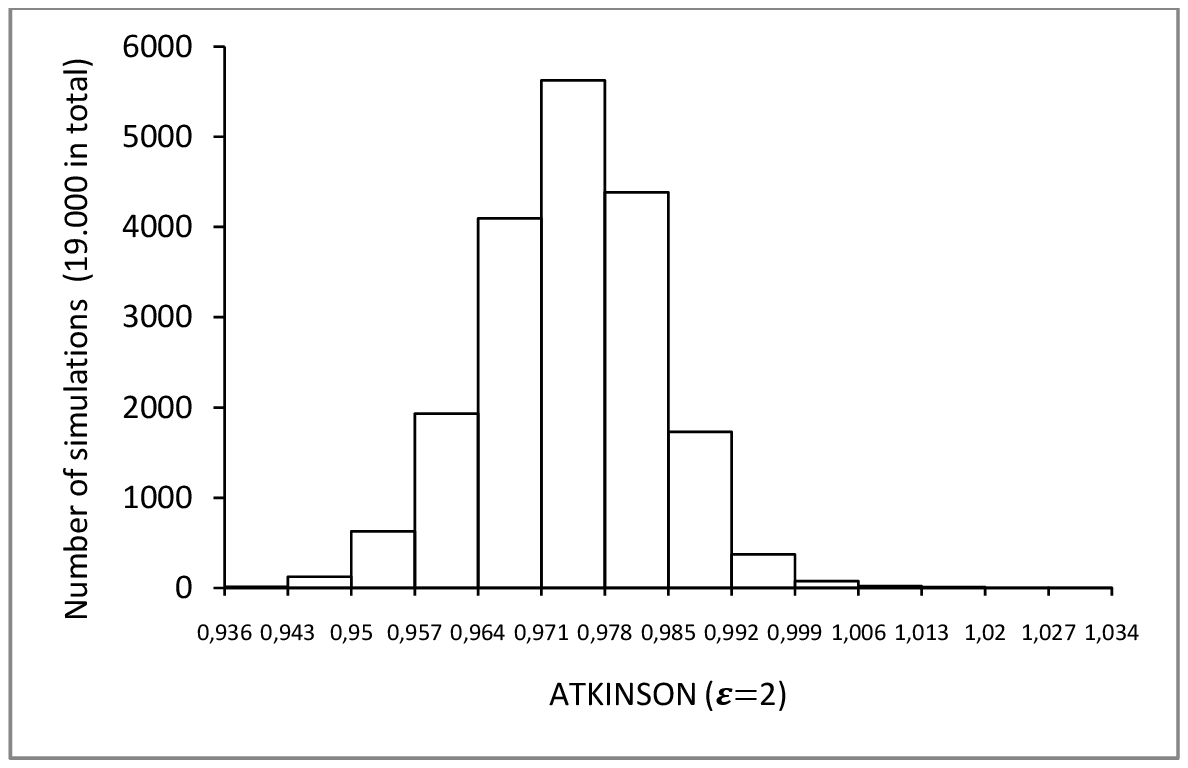}
\caption{Posterior distribution of the Gini, Theil and Atkinson indices, T=20.000, burn-in of the 1.000 first iterations}
\end{figure}

\section{Discussion}\label{s6}
Our multivariate model is similar to the model specified in Heeringa et al. (2002).
Al well we consider different models for different types of portfolios. There are still two differences.
We introduce much more information on covariates. This information is at least useful for selection issues.
We also allow for different covariance matrices in the different groups while they assume the covariance matrices are
blocks extracted from a unique matrix which here would be $5*5$. It seems that it amounts to integrating with
respect to components which are not in the portfolio as if they existed but were not observed.
We are not able to justify this choice. Also, it is not clear that the posterior of the unique latent
covariance matrix is also normal/inverse-Wishart. One relative disadvantage of our approach is that we have many
parameters. That is why we have considered only 5 "macro"-components and introduced the restriction on the parameters (RP).
It is possible to consider finer decompositions of the total wealth introducing less covariates. Note that
it is possible to avoid the introduction of group specific coefficients by specifying one multivariate
Tobit model where the residuals of the 5 latent variables are correlated. These latent variables account for both
the amount and the decision to invest in the components. We believe that this is a severe restriction. We have also
noted that, since the composition of the portfolio is observed, we do not need to model the choice mechanism and can
condition on that information.\\
\indent One of the main difficulty not treated in the paper is inconsistency implied by the fact that brackets
for the variables are not coherent with the brackets involving several variables. It allowed to detect
errors like confusion between old Francs and Euros. Concerning the final question
it turns out that it is very little informative on the top of the distribution. This is troublesome
for specification issues when we want to use only this last question. It is troublesome as well when we use
the detailed components since we have a quite poor information on the remainder. The remainder is a mixture
of luxury and durable goods and the bottom of the distribution for which the intervals are more informative
is likely to be mainly composed of durable goods.
It is thus always important, but difficult due to different selection mechanism especially due to nonresponse and
different perception of surveys, to gather information from sources exterior to the survey. It is expected for the
future French survey on Wealth to ask for the right
to use more auxiliary information from the tax declarations. Also, it is possible that
survey sampling use over sampling based on data from these tax declarations.
However, it is customary to inform
the households that their data will be matched and it sometimes increases the nonresponse rate when it is matched
with tax declarations. The threshold for the last bracket for the final question on the total wealth
was set to 450.000 \euro{} in order to be far from the threshold of 720.000 \euro{} implying eligibility
to the Solidarity Tax on Wealth and mitigate nonresponse rates to this question.

\noindent {\bf Acknowledgement.} Eric Gautier is grateful to Christian Robert for his guidance on Monte-Carlo Markov chains 
and his colleagues at the French Statistical Office INSEE among which Marie Cordier, C\'edric Houdr\'e, 
Daniel Verger and Alain Trognon for stimulating discussions on Economics of Inequalities and Econometrics.
The author also thanks the participants of the Social Statistics Seminar at INSEE and the
Econometrics Research Seminar at Yale.


\begin{thebibliography}{35}
\bibitem{A}
Arnold, S. F. (1993), "Gibbs sampling" in {\em Handbook of Statistics}, ed. C. R. R. Rao, North-Holland, 599--625.

\bibitem{CHR}
Cordier, M., Houdr\'e, C., and Rougerie, C. (2006), "Les in\'egalit\'es de patrimoine des m\'enages
entre 1992 et 2004", {\em Insee Réf\'erences}, special issue "Les revenus et le patrimoine des m\'enages", 47--58.

\bibitem{DDHFM}
Dell, F., d'Haultfoeuille, X., Février, P., and Mass\'e, E. (2002),
"Mise en {\oe}uvre de calcul de variance par lin\'earisation" in {\em Actes des Journ\'ees
de M\'ethodologie Statistique}, \verb"http://jms.insee.fr/site/index.php".

\bibitem{DSS}
Deville, J. C., S\"arndal, C.E., and Sautory, O. (1993), "Generalized Raking Procedures
in Survey Sampling," {\em Journal of the American Statistical Association}, {88}, 1013--1020.

\bibitem{DER}
Duclos, J. Y., Esteban, J., and Ray, D. (2004), "Polarization: Concepts, Measurement, Estimation,"
{\em Econometrica}, {72}, 1737--1772.

\bibitem{ER}
Esteban, J., and Ray, D. (1994), "On the Measurement of Polarization," {\em Econometrica},
{62}, 819--851.

\bibitem{EG}
Gautier, E. (2005), "El\'ements sur les m\'ecanismes de s\'election dans les enqu\^etes et sur
la non-r\'eponse non-ignorable" in {\em Actes des Journ\'ees de M\'ethodologie Statistique},
\verb"http://jms.insee.fr/site/index.php".

\bibitem{G}
Geweke, J. (2005), {\em Contemporary Bayesian Econometrics and Statistics},
Hoboken NJ, USA: Wiley.



\bibitem{HLR}
Heeringa, S. G., Little, R. J. A., and Raghunathan, T. E. (2002), "Multivariate Imputation
of Coarsened Survey Data on Household Wealth,"
in {\em Survey Nonresponse}, eds. R. M. Groves, et al., Hoboken NJ, USA: Wiley, 357--372.

\bibitem{JS}
Juster, T. F., and Smith, J. P. (1997), "Improving the Quality of Economic Data: Lessons
From the HRS and AHEAD, {\em Journal of the American Statistical Association}, {92}, 1268--1278.

\bibitem{LR}
Little, R. J. A., and Rubin, D. B. (2002), {\em Statistical Analysis with Missing Data},
2nd edition, Hoboken NJ, USA: Wiley.

\bibitem{LV}
Lollivier, S., and  Verger, D. (1988), "D'une variable discr\`ete \`a une variable
continue : la technique des résidus simulés," in {\em Mélanges économiques - Essais en
l'honneur de Edmond Malinvaud}, Economica.

\bibitem{MR}
McCulloch, R., and Rossi, P. E. (1994), "An Exact Likelihood Analysis of the
Multinomial Probit Model," {\em Journal of Econometrics}, {64}, 207--240.

\bibitem{RC}
Robert, C. P., and Casella G. (2004), {\em Monte Carlo Statistical Methods}, 2nd edition,
New York, USA: Springer-Verlag.

\bibitem{S}
Schafer, J. L. (2001), {\em Analysis of Incomplete Multivariate Data}, 2nd edition, London,
UK: Chapman \& Hall.

\bibitem{T}
Train, K. E. (2003), {\em Discrete Choice Methods with Simulation}, New York,
USA: Cambridge University Press.
\end{thebibliography}
\end{document}